\documentclass[prd,showpacs,superscriptaddress]{revtex4}
\usepackage{amsmath}
\usepackage{amssymb}
\usepackage{color}
\usepackage{graphicx}

\begin{document}

\title{Cosmic Ray Spatial Distribution and the
 Galactic/Extragalactic Transition}

\author{Paolo Lipari}

\email{paolo.lipari@roma1.infn.it}
\affiliation{INFN, Sezione Roma ``Sapienza'',
 piazzale Aldo Moro 2, 00185 Roma, Italy}

\affiliation{
 IHEP, Key Laboratory of Particle Astrophysics,
 Chinese Academy of Sciences, Beijing, China}

\affiliation{
 Tata Institute of Fundamental Research, Homi Bhabha Road, Mumbai 400005, India}

\author{Silvia Vernetto}
\email{vernetto@to.infn.it}
\affiliation{
INAF, Osservatorio Astrofisico di Torino,
 via Piero Giuria 1, 10125 Torino, Italy}

\affiliation{
 INFN, Sezione Torino,
 via P.Giuria 1, 10125 Torino, Italy}

\begin{abstract}
Determining the spatial distribution of Galactic cosmic rays (CRs)
is fundamental to understand how 
these particles propagate in interstellar space and to infer their source spectra.
The most sensitive method of studying this problem is observing the
gamma--ray and neutrino diffuse fluxes produced in the inelastic interactions
of CR protons and nuclei with interstellar gas that 
encode the energy and spatial distributions of the interacting particles.
Most theoretical models assume that the spatial and energy dependencies
of the spectra of CR protons and nuclei are factorized, so that
the CR energy spectra have the same shape at all points of the Milky Way.
However, on the base of the Fermi--LAT observations, some authors 
have tentatively inferred that particles in the inner regions of the Milky Way have 
significantly harder spectra.
Recently the ground--based telescopes Tibet--AS$\gamma$, LHAASO and HAWC
have extended the measurements of the diffuse gamma--ray flux
to much higher energies, and the IceCube neutrino telescope has
obtained evidence for a  $\nu$ flux from the Galactic disk.
These new measurements allow for new tests of the factorization hypothesis, 
and disfavor models in which the spectral shape
of CR protons and nuclei is harder in the inner part of the Galaxy.
In fact, the LHAASO data above 30~TeV are in better agreement
with the opposite hypothesis, that cosmic rays is the inner Galaxy have
softer spectra than those in the outer Galaxy.
This result, that has currently only modest significance due to
large statistical and systematic uncertainties, could be explained
by assuming that the CR confinement volume increases with energy.
An alternative explanation is that a significant fraction of cosmic rays in
the multi--PeV energy range is of extragalactic origin.
\end{abstract}

\maketitle

\section{Introduction}
The spectra of cosmic rays (CR) at the Earth are 
now known with reasonable precision over a wide range of energies. However,
much less is known about these spectra in different points in the Galaxy,
and determining their spatial and energy distributions is of great importance
for high energy astrophysics.
Information about the CR spectra in distant regions
can be obtained by observing the fluxes of
gamma--rays and neutrinos produced in the interactions of the primary cosmic ray
particles with the medium 
in which they propagate \cite{Fermi-LAT:2012edv,Lipari:2018gzn,Tibaldo:2021viq}.
The diffuse fluxes can be calculated
as the integral of the emission along the different lines of sight
and therefore encode the spatial distributions of the
interacting particles and of the targets.

It is now well understood \cite{Gabici:2019jvz,Kachelriess:2019oqu}
 that most of the cosmic rays observed
at the Earth are of Galactic origin, accelerated
in sources distributed in the volume of the Milky Way,
and confined in the Galaxy by magnetic fields 
for a time of order of millions of years, forming a ``halo'' that is
significantly larger than the thin disk that contains most
of the stars and interstellar gas. However, the
absolute normalization and spectral shape of the
cosmic ray populations in different points of the halo
are currently poorly understood, especially at high energies.

The diffuse gamma--ray flux is produced by the interactions
of both hadronic (protons and nuclei) and leptonic (electrons
and positrons) particles, while the neutrino fluxes are produced only
by hadrons.
Below few GeV the gamma--ray emission
is generated by both the hadronic and leptonic mechanisms,
but at higher energies the leptonic emission becomes negligible
because the $e^\mp$ spectra are significantly softer than
protons and nuclei spectra.
In the present work we will limit our discussion to the energy range
$E \gtrsim 1$~TeV where the gamma--ray emission
is dominated by the hadronic mechanism, and therefore 
both $\gamma$ and $\nu$ fluxes only encode information about hadronic cosmic rays. 

Most models for protons and nuclei Galactic cosmic rays
assume that their spectra in different points of the Galaxy
have the same shape, equal to what is observed at the Earth,
with an absolute normalization that changes gradually,
so that the density in the inner part of the Galaxy is larger than 
the density at large distances from the center.
This space dependence can be derived under the following assumptions:
(a) the cosmic rays accelerators in different 
regions of the Galaxy produce (on average) spectra of the same shape;
(b) the CR propagation can be modeled as diffusion, with a diffusion
coefficient that has the same rigidity dependence in all the
relevant Galactic volume;
(c) the boundaries of the diffusion volume are energy independent;
(d) the stationary solution of the diffusion equation
is a good description of the CR spectra,
at least near the solar system and in most of the Galaxy. 
However, it is possible that some (or all) of the assumptions listed above are
incorrect, and that the CR spectra have different and more complex
spatial and energy dependencies.

Most of what we know about the CR spectra in different points of the
Galaxy has been obtained by interpreting the observations of
the diffuse gamma--ray flux in the 0.1--1000~GeV energy range
by the Fermi--LAT space telescope \cite{Fermi-LAT:2012edv,Fermi-LAT:2016zaq}.
These observations are generally consistent with the
idea that the disk of the Galaxy contains
populations of cosmic rays with approximately the same
spectral shape observed at the Earth and a small density gradient.
However, several authors have suggested 
that CR spectra in the inner part of the Galaxy have harder spectra than
those at large distances from the center, and it is important
to verify these claims by extending the study to higher energies.

Recently new observations of the diffuse Galactic gamma--ray flux
in the multi--TeV to PeV energy range have been obtained
from the Tibet--AS$\gamma$ \cite{TibetASgamma:2021tpz}, 
LHAASO \cite{LHAASO:2023gne,LHAASO:2024lnz},
and HAWC \cite{HAWC:2023wdq} telescopes. In addition,
evidence for a flux of Galactic neutrinos
in roughly the same energy range has been obtained
from the IceCube detector \cite{IceCube:2023ame}.
Indications for the existence  of a flux of Galactic neutrinos
has also recently been obtained by the Baikal neutrino telescope \cite{Baikal-GVD:2024kfx}.
In the following we will discuss what these new measurements
imply for the CR spatial distributions at very high energies.

The study of of the diffuse gamma--ray and neutrino fluxes can also be a tool
for addressing a problem of central importance for cosmic ray astrophysics:
measuring the contribution of extragalactic sources to the observed CR spectra.
The CR extragalactic component is expected to be 
dominant above a transition energy $E^*$, which is not yet known, with 
different estimates suggesting values as low as few PeV ($10^{15}$~eV), 
just above the ``knee'' in the all--particle spectrum, 
and as large as $4 \times 10^{18}$~eV, where 
spectral hardening known as the ``ankle'' is observed.
At the transition energy one expects to observe
changes in the CR spectral shape and composition, but similar
changes could be also associated to the combination of CR components
of Galactic origin, produced by different classes of Milky Way sources
and having different spectral shapes and composition.
One possible method to establish the extragalactic nature of a new
CR component is to measure the energy dependence of the CR anisotropy,
however an alternative method is to infer the spatial distribution of 
the CR particles from a study of the diffuse gamma--ray an neutrino fluxes.
Indeed, if cosmic rays above a certain energy are
of extragalactic origin, one would expect their density to be 
constant or perhaps increasing towards the Galactic periphery,
while for a Galactic component one would expect
a gradient of opposite sign, with the highest density in the
inner Galaxy.
Therefore the study of the diffuse Galactic (i.e. generated in the Milky Way)
gamma--ray and neutrino fluxes can be a powerful tool to study
the CR Galactic/extra--galactic transition.

This paper is organized as follows. In the next section we discuss how 
the angular distribution of the gamma--ray and neutrino fluxes encodes the
spatial distribution of the CR spectra, and also depends 
on the distribution of interstellar gas.
In section~\ref{sec:models} we show how the Fermi--LAT observations
imply, in a model independent way, that the cosmic ray density
increases towards the Galactic center, and briefly discuss models
that claim that the CR spectral shape is position dependent.
In section~\ref{sec:high-energy-data} we discuss the recent measurements
of the Galactic gamma--rays and neutrino fluxes by
the Tibet--AS$\gamma$, LHAASO and IceCube detectors, and argue that they
disfavor models where the CR spectra are harder in the inner part of the Milky Way.
The final section contains a brief summary and outlook.

\section{The diffuse gamma--ray and neutrino fluxes}
\label{sec:diffuse_flux} 
The emission (in units [cm$^3$\,s\,GeV]$^{-1}$) of gamma--rays of
energy $E_\gamma$ from the point $\vec{x}$ due to the interactions
of CR protons with interstellar hydrogen gas can be calculated as:
\begin{equation}
 q_\gamma (E_\gamma, \vec{x}) = 
 n_{\rm ism} (\vec{x}) ~ 
\int_{E_\gamma}^\infty dE_p \; N_p(E_p, \vec{x}) ~\beta \, c \;
 \sigma_{pp} (E_p) 
 \frac{dN_{pp \to \gamma}}{dE_\gamma} (E_\gamma, E_p) ~~~~~~~~ 
\label{eq:qgamma-pp}
\end{equation}
where $n_{\rm ism} (\vec{x})$ is the density of target protons,
$N_p(E_p, \vec{x})$ is the density of CR protons
with energy $E_p$ (and velocity $\beta c$) at the point $\vec{x}$,
$\sigma_{pp} (E_p)$ is the inelastic $pp$ cross section
and $dN_{pp \to \gamma}/dE_\gamma (E_\gamma, E_p)$ is the inclusive spectrum of 
gamma--rays generated in a $pp$ inelastic collision
after the decay of all unstable particles in the final state.
The dominant source of photons is the production and decay of neutral pions
($\pi^0 \to \gamma\gamma$) with smaller contributions due to the
decays of $\eta$ and $\eta^\prime$ mesons.
Additional contributions to the gamma--ray emission
(such as $p$--helium, helium--$p$, helium--helium, and so on)
are due to interactions where the projectile and/or the target is a nucleus,
and can be calculated with integrations that have the same structure
as Eq.~(\ref{eq:qgamma-pp}). 
In the following we will consider the total 
gamma--ray emission obtained by summing over all these contributions,
assuming that CR protons and nuclei have identical 
spatial distributions, and that the interstellar medium has a constant composition.

The neutrino emission is obtained with integrations similar to
Eq.~(\ref{eq:qgamma-pp}) with obvious substitutions. In this case
the dominant $\nu$ source is the production and
chain decay of charged pions, with the decay of kaons 
contributing around 10--20\% of the emission.

The flux of gamma--rays produced by interstellar emission
observable at the Earth can be calculated by integrating
the emission along the line of sight and including 
a correction factor to take into account absorption effects \cite{Vernetto:2016alq}:
\begin{equation}
 \phi_\gamma (E, \Omega) = \frac{1}{4 \, \pi} ~
 \int_0^\infty ~dt ~q_\gamma [E, \vec{x}_\odot + t \, \hat{n}(\Omega)] ~
e^{- \tau (E, \Omega, t )}~.
\label{eq:gamma-diffuse}
\end{equation}
In this equation $\vec{x}_\odot$ the position of the solar system,
$\hat{n}(\Omega)$ is a unit vector in the direction $\Omega$,
and the factor 1/($4 \, \pi$) follows from the assumption
that the emission is isotropic.
The exponential factor is the survival probability
of gamma rays during propagation, with 
$\tau(E, \Omega, t)$ the optical depth.
The diffuse neutrino flux can be obtained with an expression very similar
to Eq.~(\ref{eq:gamma-diffuse}) with negligible absorption ($\tau=0$).

Note that gamma--rays of energy $E_\gamma$ (and neutrinos of
energy $E_\nu$) are generated by the interactions of CR
particles with a range of energy per nucleon $E_0$ that extends for approximately one decade:
($E_0/E_\gamma \sim 3$--30, $E_0/E_\nu \sim 6$--60).
The relation between the gamma--ray and neutrino energy and
the energy of the interacting nucleon is discussed in more detail in 
appendix~\ref{sec:appa}.

The crucial point in this discussion is that the
emission of gamma--rays (or neutrinos) of energy $E_{\gamma (\nu)}$
from the point $\vec{x}$ is proportional
to the product of two quantities:
the interstellar gas density $n_{\rm ism} (\vec{x})$, and 
the density of CR particles $N_{\rm cr} (E_0, \vec{x})$
with energy per nucleon $E_0$ of order ten times $E_{\gamma (\nu)}$.

Combining Eqs.~(\ref{eq:qgamma-pp})
and~(\ref{eq:gamma-diffuse})
one can immediately see that under the assumptions:
(i) the spatial and energy dependencies
of the CR spectra are factorized, so that
\begin{equation}
N_{\rm cr} (E_0, \vec{x}) \simeq N_{\rm cr} (E_0) \, f_{\rm cr} (\vec{x})
\label{eq:factorization}
\end{equation}
where $N_{\rm cr} (E_0)$ is the CR spectrum in the vicinity
of the solar system, and the (adimensional) function $f_{\rm cr} (\vec{x})$
depends only on the space coordinates
(taking value unity at the solar system position),
and (ii) the absorption effects are negligible,
then the angular and energy dependencies of the
the gamma--ray and neutrino diffuse fluxes are factorized
into the product of one function that depends only on the direction
and a second function that depends only on the energy:
\begin{equation}
 \phi_{\gamma(\nu)} (E, \Omega) = X_\phi(\Omega) \times F_{\gamma(\nu)}(E) ~.
 \label{eq:ang-fact}
\end{equation}
In Eq.~(\ref{eq:ang-fact}) the function
\begin{equation}
 F_{\gamma(\nu)} (E) \simeq \frac{1}{4 \pi} \;
 \int dE_0 ~N_{\rm cr} (E_0) \; \sigma_{pp}(E_0) \; \beta \,c \;
 \frac{dN_{\gamma (\nu)}}{dE} (E , E_0) 
 \label{eq:flux-ene} ~
\end{equation}
gives the energy distribution of the $\gamma$ ($\nu$) emission,
and is determined by the CR spectrum and composition measured at the Earth,
while the function
\begin{equation}
X_\phi(\Omega) = \int_0^\infty dt ~n_{\rm ism} [\vec{x}_\odot + t \, \hat{n}(\Omega)]
 ~f_{\rm cr} [\vec{x}_\odot + t \, \hat{n}(\Omega)] 
\label{eq:xx1}
\end{equation}
(with units cm$^{-2}$)
is the integral over the line of sight of the product $n_{\rm ism}(\vec{x}) \, f_{\rm cr} (\vec{x})$,
and describes the (energy independent) angular distribution of the
neutrino diffuse flux (at all energies), and of the gamma--ray diffuse flux 
in the energy range where the emission is dominated by the hadronic mechanism
and the effects of absorption are negligible,
that is approximately for $E_\gamma$ between 10~GeV and 100~TeV.

Testing the validity of Eq.~(\ref{eq:ang-fact}) offers a simple method
to verify (or falsify) the validity of the factorization hypothesis
for the space and energy distribution of the CR spectra
[see Eq.~(\ref{eq:factorization})] even if the spatial distribution of the target gas
and the cosmic ray energy spectra are not known.

\section{The angular distribution of the diffuse $\gamma$ and $\nu$ fluxes}
\label{sec:models}
The angular distribution of the diffuse gamma--ray and neutrino fluxes
depends on the spatial distribution of interstellar gas, that is
only known with large uncertainties.
However, the Planck satellite \cite{Planck:2016frx}
has obtained a precise map 
of the dust optical depth over the entire sky.
In the following, adopting an idea introduced by the LHAASO collaboration
\cite{LHAASO:2023gne}, we will assume that the Planck dust profile
is proportional to the quantity $X_{n}(\Omega)$ that is the
column density of interstellar gas along the same line of sight:
\begin{equation}
 X_{\rm Planck} (\Omega) \propto 
 X_{n} (\Omega) =
 \int_0^\infty dt ~n_{\rm ism} [ \vec{x}_\odot + t \, \hat{n}(\Omega)] ~.
\label{eq:xism}
\end{equation}
The quantity $X_n(\Omega)$ describes the angular distribution of the diffuse
gamma--ray and neutrino fluxes if the factorization hypothesis is valid,
and if the absolute normalization of the CR spectra is constant
in all the volume of the Galaxy where the density of interstellar gas
is not negligible.

\begin{figure}[bt]
\begin{center}
\includegraphics[width=12.9cm]{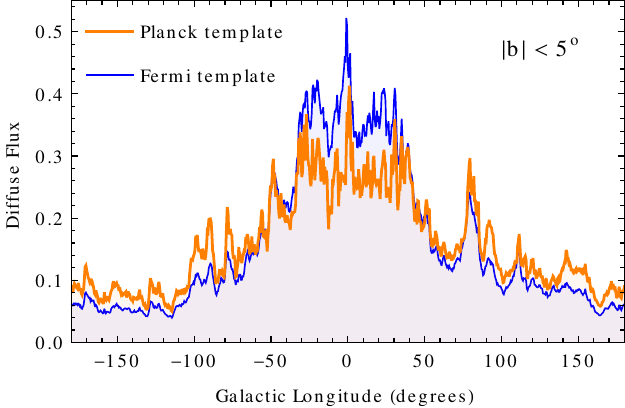}
\end{center}

\vspace{0.4cm}
\begin{center}
\includegraphics[width=12.9cm]{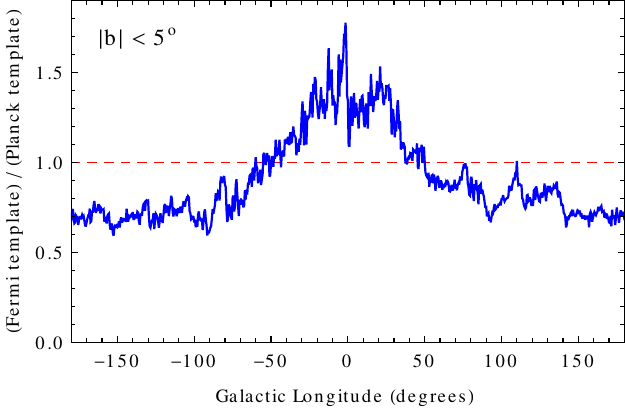}
\end{center}
\caption {\footnotesize
 The top panel shows the (normalized)
 distributions in Galactic longitude
 of the Planck--dust profile \cite{Planck:2016frx} and
 of the Fermi--LAT background model at $E_\gamma \simeq 12$~GeV
 \cite{fermi-background-model} obtained 
 integrating in the Galactic range $|b| \le 5^\circ$.
 The bottom panel shows the ratio of the two distributions.
 \label{fig:diffuse_long}}
\end{figure}

A visualization of the Planck dust profile is shown
in the top panel of Fig.~\ref{fig:diffuse_long}, that plots,
as a function of Galactic longitude, the shape of the
profile averaged in the latitude range
$|b| < b_{\rm cut} = 5^\circ$:
\begin{equation}
 \left \langle X_{\rm Planck} (\ell) \right \rangle =
 \frac{1}{4 \pi \, \sin b_{\rm cut}} \; \int_{-b_{\rm cut}}^{+b_{\rm cut}}
 db~\cos b~ X_{\rm Planck} (\ell, b) ~.
 \label{eq:long_planck}
\end{equation}
The Planck profile plotted in Fig.~\ref{fig:diffuse_long}
exibits a rich structure that encodes the spiral structure
of the Galaxy and the dishomogeneities of the interstellar
gas spatial distribution, however also some large scale features
are clearly visible: the curve is highest in directions toward
the Galactic center and lowest in directions toward the Galactic anticenter,
two broad maxima at longitudes $\ell \approx \pm 30^\circ$ indicate the existence
of a ``ring'' of gas at galactocentric radius $r \approx 4$~kpc \cite{Ferriere:2001rg}
and the peak at $\ell \approx 0$ points to a large gas density near the Galactic center.

The most accurate measurements of the diffuse gamma--ray flux
have been obtained by the Fermi--LAT space telescope
\cite{Fermi-LAT:2012edv,Fermi-LAT:2016zaq} and have been
studied in depth by several authors. The diffuse gamma--ray flux
is also the main source of background for the identification of point--like
and quasi point--like sources for a space telescope,
and the Fermi--LAT collaboration has made available
\cite{fermi-background-model} high resolution maps
(equal bins of size $0.125^\circ$ in both Galactic latitude and longitude)
of the angular dependence of the background for several energy values.
In the following we will use the Fermi--LAT background map $X_{\rm Fermi} (\Omega)$
at $E = 11.98$~GeV as a template of the angular distribution of the diffuse fluxes.
This choice of energy is motivated by the fact that at lower energies the
gamma--ray diffuse flux receives significant contributions
from the leptonic mechanism, and at higher energies the statistical
uncertainties and the possible contamination from
unresolved sources becomes larger.
One should note that the Fermi--LAT collaboration does not recommend
the use of their background maps
for studies of medium or large-scale structures in the gamma--ray sky,
but taking note of these caveats, the use of these maps
is appropriate for our purposes in the following discussion.

The shape of the
Galactic longitude distribution $\left \langle X_{\rm Fermi}(\ell) \right \rangle$
of the Fermi--LAT profile, averaged over the same 
latitude range ($|b| \le 5^\circ$) used in Eq.~(\ref{eq:long_planck}),
is shown in the top panel of Fig.~\ref{fig:diffuse_long} and compared
with the distribution of the Planck--dust template
(with both distributions normalized to unit area).
The ratio Fermi/Planck is shown in the bottom panel of the same figure.

The comparison of the Planck and Fermi--LAT templates
in Fig.~\ref{fig:diffuse_long} shows that they are remarkably similar,
with many small scale features present in both distributions;
however the figure also shows some evident differences
in the large scale structure of the two distributions:
the Fermi--LAT template is larger for small $|\ell|$
(that is for directions closer to the Galactic center)
and smaller for large $|\ell|$
(that is directions closer to the Galactic anti--center).

For a first order estimate of the size of this effect
one can look at the ratios of integrals of the two distributions
in (equal solid angle) regions around the Galactic center and anti--center.
For the Planck template one finds:
\begin{equation}
R_{\rm Planck} = \int_{|\ell| < \pi/2} d\ell ~\left \langle X_{\rm Planck} (\ell) \right \rangle
 \times \left [
 \int_{|\ell| \ge \pi/2} d\ell ~\left \langle X_{\rm Planck} (\ell) \right \rangle
 \right ]^{-1} \simeq 2.20 ~,
\label{eq:ratio-planck}
\end{equation}
while for the Fermi--LAT template the ratio
\begin{equation}
R_{\rm Fermi} = \int_{|\ell| < \pi/2} d\ell ~\left \langle X_{\rm Fermi} (\ell) \right \rangle
 \times \left [
 \int_{|\ell| \ge \pi/2} d\ell ~\left \langle X_{\rm Fermi} (\ell) \right \rangle
 \right ]^{-1} \simeq 3.35 ~~.
\label{eq:ratio-fermi}
\end{equation}
is approximately 1.5 times larger.
A simple, qualitative explanation for these results
can be easily obtained.
The Planck template [see Eq.~(\ref{eq:xism})]
maps the line integral of the column density
of the interstellar gas, while the Fermi--LAT template
maps the distribution of the product $n_{\rm ism} (\vec{x}) \times N_{\rm cr} (\vec{x})$ of the interstellar gas
and CR density.
Therefore the similarity of the two templates
indicates that cosmic rays, at least in the regions of the Milky Way
where the density of interstellar gas is not too small, 
have spectra close (in both shape and normalization) to what is observed
at the Earth. One the other hand, the differences between the two templates
indicate that the CR density (in the relevant range of energies
around $E_0 \sim 100$~GeV) is not constant in space, but is larger
toward the Galactic center and smaller toward the periphery of the Galaxy.

It is desirable to go beyond this qualitative statement, but for this task
a map of the column density of the interstellar gas
is not sufficient, and a description of the 3D distribution
of the gas is required. For this purpose 
we will use the simple, cylindrically symmetric model for the
interstellar gas density described in the paper
\cite{Lipari:2018gzn} (LV2018 in the following). 
To describe the CR radial distribution we will 
assume a form that falls exponentially with 
the (cylindrical coordinate) radius $r$,
\begin{equation}
N_{\rm cr} (r) \propto e^{-\lambda_{\rm cr} \, r} ~,
\label{eq:lambda}
\end{equation}
while the dependence the $z$--coordinate
(orthogonal to the Galactic plane) is taken 
as approximately constant in the region $|z| \lesssim 0.3$~kpc
where the density of interstellar gas is important.

\begin{figure}[bt]

\begin{center}
\includegraphics[width=12.9cm]{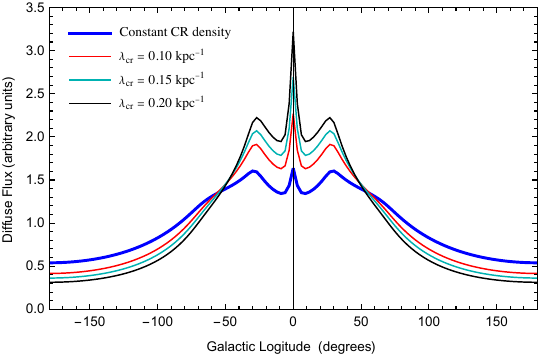}
\end{center}

\vspace{0.4cm}
\begin{center}
\includegraphics[width=12.9cm]{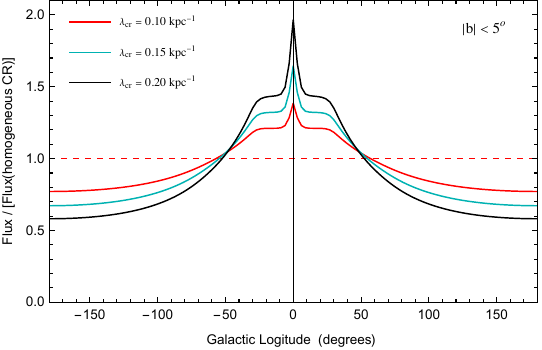}
\end{center}

\caption {\footnotesize
 The top panel shows the (normalized) longitude distribution
 of the diffuse gamma--ray and neutrino fluxes integrated over the
 latitude interval $|b| < 5^\circ$ and
 calculated assuming that the density of the interstellar gas has
 the spatial dependence described in LV2018 \cite{Lipari:2018gzn},
 and the cosmic rays spectra have energy distributions of identical shape
 in all of the Galactic disk with an absolute normalization
 that falls exponentially with the radius $r$ (in cylindrical coordinates)
 $N_{\rm cr} \propto e^{-\lambda_{\rm cr} \,r}$.
 The different lines are for a CR constant density ($\lambda_{\rm cr} = 0$)
 and for three values of $\lambda_{\rm cr}$ (0.10, 0.15 and 0.20~kpc$^{-1}$).
 The bottom panel shows the ratio between the distributions with
 $\lambda_{\rm cr} \ne 0$ and the distribution for a constant cosmic ray density.
 \label{fig:long_model}}
\end{figure}

Assuming that the CR spectral shape is identical in all points of the Galaxy,
it is straightforward to calculate the angular distribution
of the diffuse gamma--ray and neutrino fluxes
(that is independent of energy and of the shape of the CR energy spectra).
The resulting Galactic longitude distributions for this factorized model
(again averaged over the latitude range $|b| < 5^\circ$)
are shown in the top panel of Fig.~\ref{fig:long_model}
for few values of the $\lambda_{\rm cr}$ parameter
that controls the spatial distribution of the CR spectra,
including the value $\lambda_{\rm cr} = 0$ that corresponds
to a constant CR density.

The comparison of the distributions in
Figs.~\ref{fig:diffuse_long} and~\ref{fig:long_model}
shows that the simple model for the interstellar gas used in the calculation,
that does not take into account the clumpiness of its distribution
and the spiral structure of the Milky Way, 
is clearly inadequate to describe many features of the data.
However, to a first approximation, the model does reproduce the large scale
structure of the gas column density angular distribution.
The calculation with a constant CR density 
($\lambda_{\rm cr} = 0$), after integrating in the same large sky regions
used for Eqs.~(\ref{eq:ratio-planck}) and~(\ref{eq:ratio-fermi}), 
results in the flux ratio
\begin{equation}
 R_{\rm fact.mod.} (\lambda_{\rm cr} = 0) = 
 \frac{\Phi[|b| < 5^\circ, |\ell| < 90^\circ]}
 {\Phi[|b| < 5^\circ, |\ell| > 90^\circ]} \simeq 2.05 
\label{eq:r-model}
\end{equation}
that is in reasonable agreement with the ratio for the Planck template,
indicating that the space distributon of interstellar gas in the model
is an acceptable approximation.
Increasing the parameter $\lambda_{\rm cr}$, that is increasing the
gradient of the space dependence of the $\gamma$ and $\nu$ emissions,
the ratio $R_{\rm fact.mod.}(\lambda_{\rm cr})$ increases monotonically,
and the value of the ratio of the Fermi--LAT template (3.35) is reproduced for
$\lambda_{\rm cr} \simeq 0.16$~kpc$^{-1}$
(or a length scale $\lambda_{\rm cr}^{-1} \approx 6.1$~kpc)
that corresponds to a CR density near the Galactic center approximately four
times larger than the local one.

From this study one can conclude that cosmic rays with energy per nucleon
of order $E_0 \sim 100$~GeV
(that is those that produce photons of energy $E_\gamma \approx 10$~GeV)
are not uniform in the Galactic disk, but have a density
that decreases with the distance from the Galactic center,
with a gradient at the position of the solar system of order 0.16~kpc$^{-1}$.

\subsection{Possible violations of the factorization hypothesis in the Fermi--LAT data}
\label{sec:factor-violations}
Several authors
\cite{Gaggero:2014xla,Gaggero:2015xza,Yang:2016jda,Fermi-LAT:2016zaq}
analysing the data obtained by Fermi--LAT on the diffuse
gamma--ray flux have suggested that the emission from the 
innermost region of the Galactic disk is significantly harder
than the local (near the solar system) emission, 
while the emission from points far from the Galactic center is softer.

Such a spatial dependence of the CR spectral shape effect
could be caused by the fact that 
different regions of the Galaxy contain different CR accelerators,
with sources in the inner Galaxy producing harder
spectra than those in the outer Galaxy. An alternative explanation is that
this is a propagation effect, with cosmic rays in different Galactic regions
having different rates of escape. This idea is implemented in the
so called KRA$_\gamma$ models \cite{Gaggero:2014xla,Gaggero:2015xza}
that are part of the DRAGON code for the transport of cosmic rays in the Milky Way.
In these models the rigidity dependence of the diffusion coefficient 
is a function of the space coordinates, and this 
results in harder CR spectra in the inner Galaxy.

In a previous work \cite{Lipari:2018gzn} we have constructed 
a non--factorized model for the diffuse gamma--ray emission 
where its spectral shape becomes progressively softer with the galactocentric radius $r$.
In the following we will use this ``gradient model'',
where the spatial and energy dependencies of the gamma--ray emission are similar to
those of the KRA$_\gamma$ models \cite{Gaggero:2014xla,Gaggero:2015xza},
to calculate predictions that can be compared to the high energy
observations of the diffuse gamma--ray and neutrino fluxes.

\begin{figure}[bt]

\begin{center}
\includegraphics[width=12.9cm]{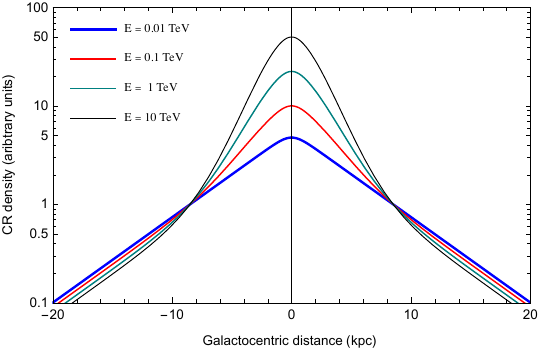}
\end{center}

\vspace{0.4cm}
\begin{center}
\includegraphics[width=12.9cm]{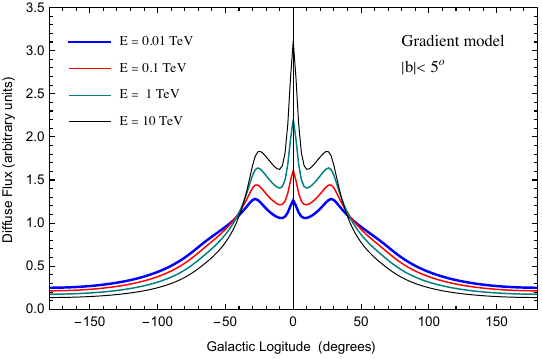}
\end{center}

\caption {\footnotesize This figure shows distributions calculated for
the non--factorized gradient model [see Eq.~(\ref{eq:q-gradient})].
The top panel plots,
as a function of the galactocentric radius $r$ (in cylindrical coordinates),
the ratio $q_\gamma (E,\vec{x})/n_{\rm ism} (\vec{x})$ between the emission of
gamma--rays of different energies ($E = 0.01$, 0.1, 1 and 10~TeV)
and the interstellar gas density at the point $\vec{x}$.
The ratio is shown normalized to unity at the position of the solar system,
and is proportional to the density of the primary cosmic rays
that generate the gamma--rays of the energy considered.
The bottom panel shows the longitude distributions 
of the diffuse gamma--ray flux,
averaged over the Galactic latitude interval $|b| < 5^\circ$
and normalized to equal area, for the same values of the energy.
 \label{fig:space_cr_gradient}}
\end{figure}

In the ``gradient model'' 
the gamma--ray emission is described by the form:
\begin{equation}
 q_\gamma (E, \vec{x}) = q_\gamma (E, \vec{x}_\odot) ~ 
 \frac{n_{\rm ism} (\vec{x})}{n_{\rm ism} (\vec{x}_\odot)} ~
 \left ( \frac{E}{E_{\gamma,0}} \right )^{-\Delta \alpha (r)} 
\label{eq:q-gradient}
\end{equation}
with $q_\gamma (E, \vec{x}_\odot)$ the local emission
(at the position of the solar system),
that is rescaled with the value $n_{\rm ism} (\vec{x})$
of interstellar gas density at the position $\vec{x}$,
and distorted by the factor $(E/E_{\gamma,0})^{-\Delta \alpha(r)}$
with $E_{\gamma,0} = 12$~GeV. The exponent $\Delta \alpha(r)$ is a function
of the (cylindrical coordinate) radius $r$ and 
takes values between $+0.35$ at the Galactic center
(where the emission is hardest) and $-0.05$ for large $r$
(where the emission is softest) vanishing at $r = r_\odot$.
The energy distribution of the local emission $q_\gamma (E, \vec{x}_\odot)$
(in the high energy range discussed here)
is calculated using the CR spectra
and composition of the Gaisser, Stanev and Tilav (GST) model \cite{Gaisser:2013bla}
and the Sibyll--2.1 code \cite{Ahn:2009wx} to model hadronic interactions.
The space distributions of the emission predicted by this model
for different values of the gamma--ray energy:
($E_\gamma = 0.01$, 1, and 10~TeV) are shown in the top panel of
Fig.~\ref{fig:space_cr_gradient},
where one can see that for larger $E_\gamma$ the 
emission from the inner part of the Galaxy becomes increasingly more important.
The bottom panel in Fig.~\ref{fig:space_cr_gradient} shows the longitude
distributions, averaged in the latitude interval $|b| < 5^\circ$,
of the gamma--ray diffuse flux predicted by the model
for different values of the energy. The longitude distributions have
an energy dependent shape, and the ratio between the fluxes
in the directions of the Galactic center and anti--center becomes
larger with increasing energy.

\section{High Energy measurements of the gamma--ray diffuse flux}
\label{sec:high-energy-data}
The indications for the non--factorization of the cosmic ray spectra
discussed in section~\ref{sec:factor-violations} 
have been obtained by interpreting 
Fermi--LAT data on the diffuse gamma--ray flux in an energy range 
than spans approximately two decades (10--10$^3$~GeV).
More recently the ground based telescopes 
Tibet--AS$\gamma$ \cite{TibetASgamma:2021tpz},
LHAASO \cite{LHAASO:2023gne,LHAASO:2024lnz}
and HAWC \cite{HAWC:2023wdq} telescopes 
have obtained measurements of the gamma--ray diffuse flux
at much higher energies, in the 1--$10^3$~TeV range.
The interpretation of these observations requires to separate 
the (truly diffusive) component of the signal produced by CR interactions
with the interstellar medium from the contributions
due to the unresolved, discrete sources, a question that has been
discussed by several authors
\cite{Yan:2023hpt,Fang:2023ffx,Lipari:2024pzo,Lipari:2025nta,Vecchiotti:2024kkz,DeLaTorreLuque:2025zsv}.
In the following we will assume that the unresolved sources contribution
is only subdominant, and discuss how 
the extension to higher energies of the diffuse gamma--ray flux 
allows to obtain information about the energy dependence of the
CR spatial distribution.

\begin{figure}[bt]
\includegraphics[width=8.0cm]{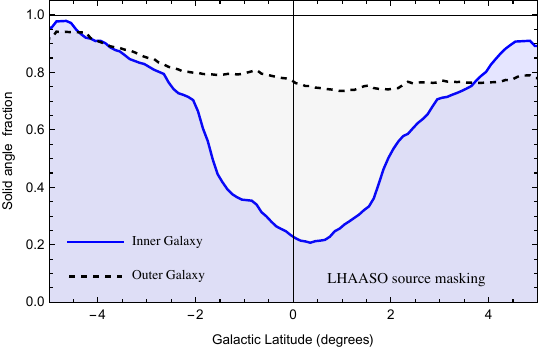}
~~~~ \includegraphics[width=8.0cm]{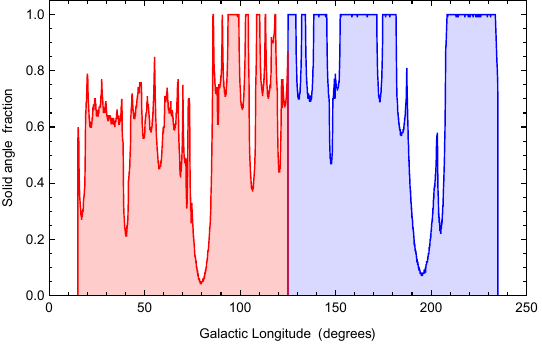}
\caption {\footnotesize
 Effect of the masking of known gamma--ray sources used by the LHAASO
 telescope in reducing the solid angle available for measurement of the
 diffuse gamma--ray flux.
 The left panel shows the reduction plotted as a function of
 Galactic latitude for the Inner--Galaxy region
 (longitudes $15^\circ \le \ell \le 125^\circ$) and Outer--Galaxy region
 (longitudes $125^\circ \le \ell \le 235^\circ$).
 The right panel shows the reduction plotted as a function of
 Galactic longitude after integration in the latitude range
 $ |b| \le 5^\circ$. The different shadings show the Inner and Outer--Galaxy
 regions.
 \label{fig:mask_lhaaso}}
\end{figure}

A ground based telescope can measure the gamma--ray flux
only in a limited sky region.
The published HAWC \cite{HAWC:2023wdq} measurements of the diffuse gamma--ray flux
cover only a small fraction of the celestial sphere,
and will not be used in the following discussion. The Tibet--AS$\gamma$ and
LHAASO telescopes have released measurements of the diffuse gamma--ray flux
in two large regions of the Galactic disk
that are described in Table~\ref{tab:gamma-diffuse-regions}.
\begin{table}[hbt]
 \caption{\footnotesize
 Sky regions where measurements of the diffuse gamma--ray flux
 at very high energy have been obtained.
\label{tab:gamma-diffuse-regions}}
 \renewcommand{\arraystretch}{1.4}

 \vspace{0.20cm}
 \begin{tabular}{ | l | l || c | c | c |}
 \hline
Telescope & Region & Latitude & Longitude & Non--masked fraction \\
\hline %
Tibet--AS$\gamma$ & Inner Galaxy & $|b| < 5^\circ$ & $25^\circ\le \ell \le 100^\circ$
 & 1 \\
 Tibet--AS$\gamma$ & Outer Galaxy & $|b| < 5^\circ$ & $50^\circ\le \ell \le 200^\circ$
 & 1 \\
\hline
LHAASO & Inner Galaxy & $|b| < 5^\circ$ & $15^\circ\le \ell \le 125^\circ$ & 0.616 \\
LHAASO & Outer Galaxy & $|b| < 5^\circ$ & $125^\circ\le \ell \le 235^\circ$ & 0.801 \\
\hline
 \end{tabular}
\end{table}
One should note that the two regions of the measurements
by the Tibet--AS$\gamma$ telescope have a significant overlap, and that the
LHAASO measurements are performed in reduced angular regions
due to masking of directions close to known gamma--ray sources.
This corresponds to the exclusion of 
approximately 38\% of the Inner--Galaxy region and 20\% of the Outer--Galaxy region
(2D maps of the LHAASO sky masking can be seen in \cite{LHAASO:2023gne}).
Figure~\ref{fig:mask_lhaaso} shows the effect of the masking in reducing the
solid angle of the LHAASO observations as a function of Galactic
latitude (top panel) and longitude (bottom panel). One can see that the masking
mostly excludes points in the sky that have small $|b|$ and small $\ell$, that is
directions that look ``inside'' the Galactic disk and toward the Galactic center.
These are in fact the most interesting directions for a study of
the CR space distribution, because they traverse more distant regions of the
Milky Way. Therefore, the masking has the effect to significantly
reduce the sensitivity of the LHAASO observations in the study of 
the energy dependence of the CR spatial distributions.

As discussed above, 
if the factorization hypothesis of Eq.~(\ref{eq:factorization}) is valid,
and absorption effects are negligible, then the ratio:
$\langle \phi_{\gamma} (E, \Delta \Omega_1) \rangle /
\langle \phi_{\gamma} (E, \Delta \Omega_2) \rangle$
between the diffuse fluxes at the same energy $E$, but 
averaged over two different sky regions
$\Delta \Omega_1$ and $\Delta \Omega_2$ should be energy independent.
Conversely, an energy dependence of the ratio implies
that the CR spectra have different spatial distributions at different energies.

\begin{figure}[bt]
\begin{center}
\includegraphics[width=12.9cm]{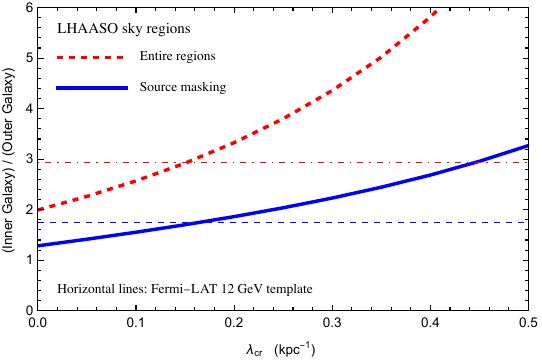}
\end{center}
\caption {\footnotesize
 Ratio of the average diffuse gamma--ray fluxes in the Inner--Galaxy
 \{$|b| < 5^\circ$, $15^\circ\le \ell \le 125^\circ$\}
and Outer--Galaxy
\{$|b| < 5^\circ$, $125^\circ\le \ell \le 235^\circ$\} regions
used by the LHAASO telescope. The average fluxes are calculated
in two ways: integrating over the entire regions
and integrating in the reduced solid angles obtained
masking kwown sources. The diffuse flux calculations are performed
using the model of interstellar gas described in LV2018 \cite{Lipari:2018gzn}
and cosmic ray spectra that have energy distributions of the same shape
in all the Galactic disk and an absolute normalization
that falls exponentially with the radius $r$:
$N_{\rm cr} (E, \vec{x}) \propto e^{-\lambda_{\rm cr} \, r}$.
The results are plotted as a function of the parameter $\lambda_{\rm cr}$.
The horizontal lines show the value of the ratio estimated assuming
that the gamma--ray flux has the angular distribution of the
Fermi--LAT 12~GeV template.
 \label{fig:ratio_space_slope}}
\end{figure}

To illustrate the sensitivity of this method to the CR space
distributions, in Fig.~\ref{fig:ratio_space_slope} we show the ratio
$\langle \phi_{\rm inner}\rangle / \langle \phi_{\rm outer} \rangle$ between
the average gamma--ray fluxes in the two LHAASO sky regions
calculated in the simple factorized model discussed in the previous
session, plotted as a function of the parameter $\lambda_{\rm cr}$.
One can see that the ratio increases monotonically with $\lambda_{\rm cr}$,
because a larger value of the parameter
implies a larger (smaller) CR density in inner (outer) part of the Galaxy.
In the figure two different lines show the inner/outer ratio
calculated averaging the fluxes 
over the reduced solid angle regions obtained by masking the known sources
(as done in the LHAASO measurement) and by averaging over the entire regions.
The dependence of the inner/outer ratio on the shape of the CR spatial
distribution is significantly larger in the second case, when the 
fluxes are averaged over all lines of sight, including those that
traverse the inner part of the Galactic disk.

\subsection{The LHAASO telescope observations}

\begin{figure}[bt]
\begin{center}
\includegraphics[width=12.9cm]{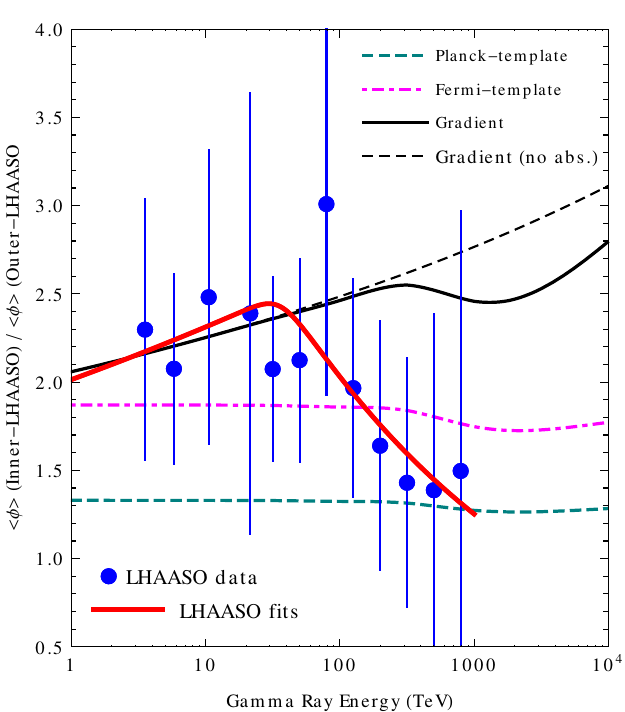}
\end{center}
\caption {\footnotesize
 Ratio of the average gamma--ray diffuse fluxes
 in the Inner--Galaxy
 $\{|b| < 5^\circ, 15^\circ \ell \le 125\}$
 and Outer--Galaxy
 $\{|b| < 5^\circ, 125^\circ \ell \le 235\}$ regions
 observed by the LHAASO telescope \cite{LHAASO:2023gne}
 plotted as a function of energy. The average fluxes
 are for the reduced angular regions obtained masking
 known sources.
 The points are the measurements of the LHAASO telescope,
 and the thick red line is the ratio of the fits to the
 measured average fluxes in the two regions obtained by the
 LHAASO collaboration \cite{LHAASO:2023gne}.
 The other lines show the ratio calculated with different models
 for the angular distribution of the gamma--ray diffuse flux:
 the Planck--dust template, the Fermi--LAT template and the
 non--factorized model of LV--2018. For the last model the ratio calculated
 without taking into account absorption is also shown.
 For the other two models the unabsorbed ratio is a constant
 with the same value obtained at low energy.
 \label{fig:ratio_mask}}
\end{figure}

The ratio $R_{\rm Lhaaso} (E) =
\langle \phi_{\rm inner} (E) \rangle /
\langle \phi_{\rm outer} (E) \rangle$
of the LHAASO measurements \cite{LHAASO:2024lnz}
of the diffuse gamma--ray flux in the inner and outer
Galaxy sky regions (described in Table~\ref{tab:gamma-diffuse-regions})
are shown in Fig.~\ref{fig:ratio_mask}
as points with error bars.
The LHAASO collaboration has also fitted the flux measurements 
with a broken power--law form, 
and the ratio $\phi_{\rm inner}^{\rm fit} (E)/\phi_{\rm outer}^{\rm fit}(E)$
of the two best fits to the spectra is also shown in the figure as a thick line.

The other lines in Fig.~\ref{fig:ratio_mask} corresponds to the ratio
calculated for different models.
Two of the models assume the validity of the factorization
hypothesis of Eq.~(\ref{eq:factorization}) and therefore predict 
a constant ratio in the energy range ($E \lesssim 100$~TeV) where
absorption effects are negligible. These constant values are 
1.33 and 1.87 
for the Planck and Fermi--LAT template, respectively.
As discussed above, the Planck template, that corresponds to a constant
CR density, predicts a smaller flux ratio.

At higher energies the inner/outer flux ratio predicted by a factorized model decreases,
because the absorption effects are on average larger for
for lines of sights in the inner Galaxy sky region, 
when the flux receives contributions from points at larger distance
from the Earth. 
The calculation of the absorption effects requires a 3D model for the
spatial distributions of cosmic rays and interstellar gas,
and for this purpose we have used the models described in section~\ref{sec:diffuse_flux}.
The absorption effects are a modest correction
that is largest (approximately 6\% and 9\% for the Planck and Fermi--LAT
templates, respectively)
at $E_\gamma \simeq 2.2$~PeV, the energy where the absorption length is shortest
because of the large cross section
for pair production interactions ($\gamma\gamma \to e^+e^-$)
with the photons of the Cosmic Microwave Background.

The third prediction of the flux ratio in Fig.~\ref{fig:ratio_mask}
has been calculated for the non--factorized ``gradient model''
introduced in section~\ref{sec:factor-violations},
and is shown by two lines which give the ratio obtained 
by taking into account or neglecting the absorption effects.
This is constructed to obtain the same inner/outer flux ratio 
of the Fermi--LAT template ($R \simeq 1.87$) at $E \simeq 12$~GeV,
but the predicted ratio grows with energy, reflecting the
fact the CR spectra are harder
in the inner part of the Galaxy, taking the values 
$R \simeq 2.1$ at $E_\gamma \simeq 1$~TeV and
$R\simeq 2.5$ (2.8 without absorption) at $E_\gamma \simeq 10^3$~TeV.
The absorption effects are a little larger (with a maximum of order 13\%)
than those calculated for the factorized models, because the contributions
to the flux of the inner (and more distant) regions of the Galaxy are larger. 

The comparison of the LHAASO measurements of the inner/outer flux ratio
with the predictions of the three models shown in Fig.~\ref{fig:ratio_mask}
can be used to infer the properties and energy dependence of the
CR spatial distribution, however the data points 
have very large errors, and the differences between the model predictions 
are rather modest, and therefore it is not possible to reach firm conclusions.
However, this comparison does give some intriguing indications.
Combining the LHAASO measurements 
in three energy intervals: ($E \le 30$~TeV), ($30 < E < 200$~TeV) and
($E \ge 200$~TeV) each containing four data points, 
one finds the average values of the flux ratio:
$\langle R_{\rm Lhaaso} \rangle \simeq 2.31 \pm 0.44$,
$2.29 \pm 0.27$ and $1.49 \pm 0.51$ respectively.
These results, taking into account the large errors,
are consistent with the factorization hypothesis and 
an energy independent ratio of order $\approx 2.0$,
approximately equal to what is predicted by assuming
that the angular distributions of the high energy gamma--ray flux remain
equal to what has been observed in the GeV energy range by Fermi--LAT.
The non--factorized gradient model predicts for the highest energy interval a
ratio 2.51, a little larger (a 2$\sigma$ deviation) than the measurement.
It is also interesting to note that the measured flux ratio becomes 
smaller at the highest energies.
This result can be also described by using the published fits to the measured fluxes,
as the ratio of the fits takes at high energy the asymptotic power--law form:
\begin{equation}
 \left . \frac{\phi_{\rm inner}^{\rm fit}(E)}{\phi_{\rm outer}^{\rm fit}(E)}
 \right |_{E \gtrsim 30~{\rm TeV}} \propto E^{-0.21 \pm 0.13}
\label{eq:fit1}
\end{equation}
(the error on the exponent in Eq.~(\ref{eq:fit1}) includes only statistical uncertainties).
Models where very high energy (PeV range)
cosmic rays have harder spectra in the inner part of the Galaxy
predict an inner/outer flux ratio that increases with energy,
and are therefore disfavored by these results.
In fact, a flux ratio that decreases with energy would imply
the opposite, that is softer CR spectra in the inner Galactic region.
This energy dependence of the CR spatial distributions could be the effect of
cosmic ray confinement volume that becomes larger
at higher energies.

\begin{figure}[bt]
\begin{center}
\includegraphics[width=12.9cm]{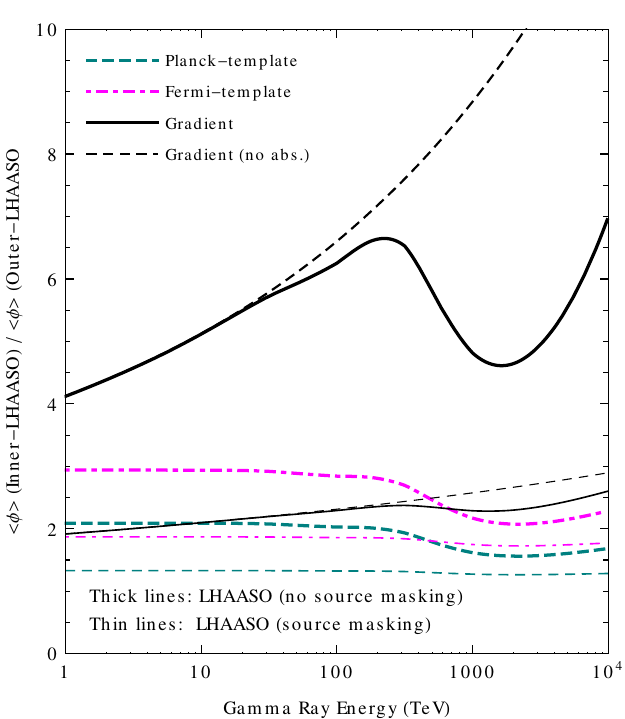}
\end{center}
\caption {\footnotesize Ratio of the average gamma--ray fluxes in the
 inner Galaxy and Outer Galaxy regions observed by the LHAASO detector
 according to three models: the Planck dust template,
 the Fermi--LAT 12~GeV template, and the Gradient model of LV2018.
 The thick lines show the ratio for fluxes averaged over the
 entire sky regions. The thin lines show the ratio for fluxes
 averaged over the reduced solid angle obtained by masking directions
 around known sources. The difference between the models is much larger
 when the diffuse fluxes are measured in all solid angle.
 \label{fig:ratio_lhaaso}}
\end{figure}

The LHAASO measurements and the model predictions of the flux ratio
discussed above have been obtained by averaging the gamma--ray fluxes over
reduced sky regions that exclude directions around known gamma--ray sources.
This source masking excludes preferentially
(see Fig.~\ref{fig:mask_lhaaso}) lines of sight at small
latitude and close to the Galactic center, that are those
most sensitive to the CR spatial distribution.
Measuring the diffuse flux also in the directions
that have been masked in the published LHAASO measurements
would significantly increase the sensitivity
of these studies to the cosmic ray spatial distribution.
This is illustrated in Fig.~\ref{fig:ratio_lhaaso} that shows
the predictions of the flux ratio
for the same models 
and the same sky regions as discussed above, but including also the
directions that were previously excluded due to source masking.
In this case the differences between the predictions
is much larger, and accordingly
the sensitivity to the CR spatial distribution is enhanced.
Such an extension of the diffuse flux measurements is therefore
in principle very desirable,
however this requires the subtraction of the source contributions, 
a procedure that can be difficult and problematic.

\subsection{The Tibet--AS$\gamma$ telescope observations}
\begin{figure}[bt]
\begin{center}
\includegraphics[width=12.9cm]{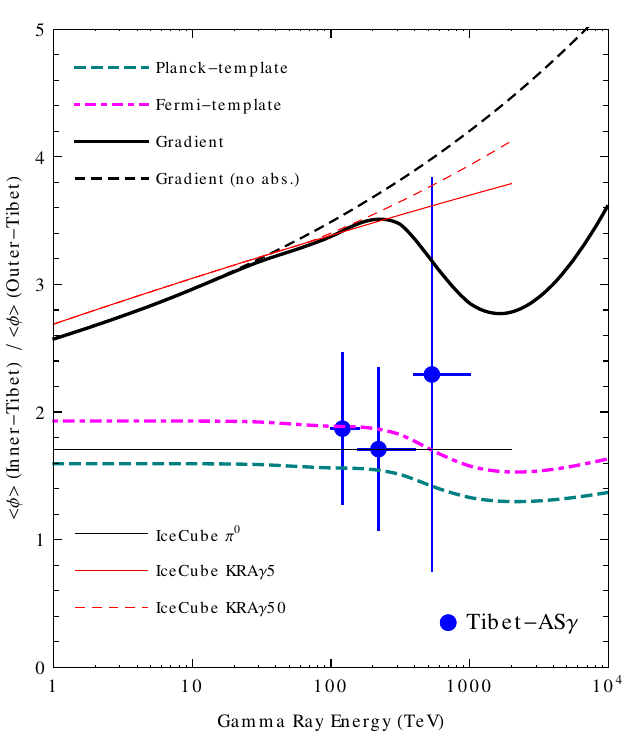}
\end{center}
\caption {\footnotesize
 Ratio of the average gamma--ray diffuse fluxes
 in the Inner--Galaxy
 $\{|b| < 5^\circ, 25^\circ \ell \le 100\}$
 and Outer--Galaxy $\{|b| < 5^\circ, 50^\circ \ell \le 200\}$
 regions observed by the Tibet--AS$\gamma$ telescope
 \cite{TibetASgamma:2021tpz}
 plotted as a function of energy.
 The points show the ratio of the fluxes measured 
 by the Tibet--AS$\gamma$ telescope.
 The thick lines show the ratio calculated with three models
 for the angular distribution of the gamma--ray diffuse flux:
 the Planck--dust template, the Fermi--LAT template (for $E_\gamma = 12$~GeV) and the gradient model. For the last model we also show the ratio
 calculated neglecting absorption effects
 (for the Planck and Fermi--LAT models, the no--absorption result is
 a constant, with the same value obtained at low energy).
 The thin lines show the ratio for the models used
 by IceCube as templates for the neutrinos
 diffuse flux ($\pi^0$, KRA$_{\gamma}^{5}$ and KRA$_{\gamma}^{50}$ models).
 \label{fig:ratio_tibet}}
\end{figure}

The Tibet--AS$\gamma$ telescope 
has published \cite{TibetASgamma:2021tpz}
measurements of the average diffuse gamma--ray flux
in three energy intervals (100--158~TeV, 158--398~TeV and 398--1000~TeV)
for the two regions of the Galactic disk described
in Table~\ref{tab:gamma-diffuse-regions}.
The ratios of the flux measurements in the two sky regions 
are shown in Fig.~\ref{fig:ratio_tibet} as points with error bars
and have values $1.9\pm 0.6$, $1.7\pm 0.6$ and $2.3 \pm 1.5$,
that, taking into account the very large errors,
are consistent with an energy independent, constant value.

The Tibet--AS$\gamma$ results on the flux ratio
are shown together with the theoretical
predictions calculated for the same three models discussed above.
The two factorized models
for $E \lesssim 100$~TeV (where gamma--ray absorption is negligible)
predict constant values of the flux ratio: $R \approx 1.59$ and 1.98,
by assuming the angular distributions of
the Planck and Fermi--LAT templates, respectively.
At higher energies the absorption effects generate a small decrease 
of the flux ratio with maximum reduction
($-18$\% and $-21$\% for the Planck and Fermi--LAT templates, respectively)
at $E \approx 2.2$~PeV.

For the gradient model, where the energy spectra are harder in the
inner part of the Galaxy, the predicted flux ratio is larger,
taking the value $R\simeq 2.6$ at 1~TeV, and grows with energy,
reaching a maximum value ($R \simeq 3.5$) at $E \sim 250$~TeV.
If absorption effects are neglected the calculated
flux ratio would grow monotonically with energy,
but including these effects this growth is arrested, 
and the ratio starts to decrease, reaches a minimum 
at $E \approx 3$~PeV, and then grows again,
as absorption effects become smaller at very high energies.

The very large errors of the measurements do not allow to
firm conclusions from a comparison of the 
Tibet--AS$\gamma$ telescope observations with the predictions.
However, also in this case, the meaurements of the flux ratio
are smaller than the predictions of models (such as the ``gradient model'')
that assume that cosmic rays have harder energy spectra in the inner part of the Galaxy,
and are consistent with an energy independent
flux ratio, equal to what is measured at lower energies
by Fermi--LAT.

\subsection{The IceCube measurement of Galactic neutrinos}
\label{sec:icecube}
The IceCube neutrino telescope 
has recently reported \cite{IceCube:2023ame} evidence
for the existence of a flux of neutrinos from the Galactic disk.
The IceCube  evidence has been obtained by comparing the observations
to a background--only hypothesis and to the predictions of three templates
for the energy and angular distributions of the diffuse $\nu$ flux.
In one template (the so called $\pi^0$ template)
the neutrino angular and energy distributions
are factorized, with the angular distribution 
equal to the one observed by Fermi--LAT for gamma--rays at $E \sim 10$~GeV,
while the energy distributions has a simple power--law form $\propto E^{-2.7}$.
This model is similar to the Fermi--LAT template discussed above in this paper.
The other two templates are the KRA$_\gamma^5$
and KRA$_\gamma^{50}$ models \cite{Gaggero:2014xla,Gaggero:2015xza},
already introduced in section~\ref{sec:factor-violations}, that
are constructed by assuming that the cosmic ray spectra, and therefore the
$\gamma$ and $\nu$ emission spectra, are harder in the central region of the Galaxy,
with predictions similar to those of the gradient model discussed in this paper.

Comparing the data and the predictions of the templates,
with the absolute normalization of the total diffuse flux
as the only free parameter, the IceCube collaboration has obtained
evidence for the existence of a Galactic neutrino flux
with a significance of 4.71, 4.36 and 3.96 $\sigma$'s
for the $\pi^0$, KRA$_\gamma^5$ and KRA$_\gamma^{50}$ templates, respectively.
This method is sufficient to establish the existence
of a Galactic neutrino flux, but gives only a limited information
about its angular and energy distributions. All three
templates, even if they significantly differ from each other,
are consistent with the observations.

A clear illustration of the differences between the predictions of the
three templates is contained
in figure~S8 of the Supplemental material for
the IceCube paper \cite{IceCube:2023ame}.
This figure shows the energy spectra 
obtained by averaging over the inner--Galaxy and outer--Galaxy sky regions
used by the Tibet--AS$\gamma$ telescope 
the diffuse $\nu$ fluxes predicted by the three templates, 
all plotted with the absolute normalization of the best fit.
The figure also shows the neutrino fluxes inferred
from the measurements of the Tibet--AS$\gamma$ telescope \cite{TibetASgamma:2021tpz}
by transforming the observed gamma--ray fluxes into neutrino fluxes
by using the simple rule:
\begin{equation}
\phi_\nu (E_\nu) \simeq 6 \, \phi_\gamma ( 2 E_\nu) ~.
\label{eq:nuflux}
\end{equation}
This rule can be derived by assuming that
(i)  both fluxes are generated by
the hadronic mechanism with the decay of charged (neutral) pions
as the dominant source for the production of $\nu$ ($\gamma$)
and  (ii) that the pion production ratio is $\pi^\pm/\pi^0 \simeq 2$,
and also  noting that each photon
carries on average half of the parent energy, while each of the three neutrinos
in a $\pi^\pm$ chain decay carries approximately one--quarter of the parent energy.

Inspecting figure S8 in the IceCube paper \cite{IceCube:2023ame}
one can note some interesting points.
(a) The $\pi^0$ template obviously predicts
spectra of equal shape ($\propto E^{-2.7}$) in both inner and outer--Galaxy sky regions,
while the two KRA$_\gamma$ models predict much harder $\nu$ spectra
in the inner--Galaxy region.
(b) All three templates, with the absolute normalizations obtained from 
best fits estimated globally (over the entire sky), give neutrino spectra
of approximately the same size and similar to 
the $\nu$ flux inferred from the Tibet--AS$\gamma$ data using
Eq.~(\ref{eq:nuflux}). However, the templates have different angular
distributions, and the predicted inner/outer ratio is much larger
for the KRA$_\gamma$ models. Accordingly, in the outer--Galaxy region only the
the $\pi^0$ template predicts a neutrino flux 
consistent with the Tibet--AS$\gamma$ data,
while the flux predicted by the two KRA$_\gamma$ models are smaller
by a factor $\approx 2.5$.

Essentially the same results are illustrated, in a different form,
in Fig.~\ref{fig:ratio_tibet}, that shows (as thin lines) the energy dependence of 
the ratio of the gamma--ray fluxes in the inner and outer--Galaxy regions
predicted by the three IceCube templates.
These fluxes are calculated by reading the neutrino fluxes 
from the IceCube publication 
and converting them into gamma--ray fluxes using the rule of
Eq.~(\ref{eq:nuflux}), a procedure that gives results that do not
take into account the effects of gamma--ray absorption.
The $\pi^0$ template predicts a constant ratio with a value close to the
measurement of the Tibet--AS$\gamma$ telescope 
and to what is obtained using our Fermi--LAT template.
On the other hand the KRA$_\gamma^{5}$ and KRA$_\gamma^{50}$ templates predict
a larger flux ratio that increases with energy, with results very similar to the 
predictions of our gradient model.

This discussion indicates that if the Galactic neutrinos observed by
IceCube and the gamma--rays observed by the Tibet--AS$\gamma$ telescope
have the same origin and are generated by interstellar emission,
then the $\pi^\circ$ template is a better description of the neutrino fluxes
than the KRA$_\gamma$ models, because it describes correctly
the spectra in both the inner and outer Galaxy sky regions.
The underprediction for the KRA$_\gamma$ models of the flux in
the outer--Galaxy region does not significantly affect
the quality of the fits because the neutrino signal in this angular region is
below the detector sensitivity.
This tentative conclusion favors models where the spatial distribution of
cosmic rays at very high energy keeps the same shape (or becomes broader)
than what has been observed in the GeV energy range by Fermi--LAT,
and disfavors models where cosmic rays have harder spectra in the central
region of the Galaxy.

\section{Summary and outlook}
\label{sec:summary}
The recent extension of the gamma--ray observations
to very high energies by LHAASO, Tibet--AS$\gamma$ and HAWC,
and the first measurements of Galactic neutrinos by IceCube
give valuable information about the cosmic ray spatial distribution at
very high energies, a question of fundamental importance
for high energy astrophysics with important and broad implications.
However, the interpretation of the observations
to infer the CR distributions is a non--trivial problem.
An important source of uncertainty is the modeling of the 
spatial distribution of interstellar gas, the target for the cosmic ray
interactions where neutrinos and gamma--rays are produced.

In this work we have used a simple approach that allows
to determine whether the cosmic ray spatial distribution 
is energy dependent without any knowledge or assumption about
the interstellar gas distribution, and about the 
shape of the cosmic ray spectra.
The simple idea is to compare the shapes of the angular distributions 
of the diffuse $\gamma$ and $\nu$ fluxes at different energies.
If this shape remains constant,
one can infer that cosmic rays of different energies
have the same spatial distribution, or equivalently 
that cosmic rays have the same spectral shape
in all points of the Galaxy where the $\gamma$ and $\nu$ emission is significant.
Conversely if, increasing the energy, the angular distributions
of the diffuse fluxes become more (less) peaked in directions toward
the Galactic center, one can conclude
that, increasing the energy, the ratio of the cosmic ray densities 
in the inner and outer part of the Milky Way becomes larger (smaller).

The observations of the Tibet--AS$\gamma$ and LHAASO gamma--ray telescopes
are consistent with the hypothesis that the angular distribution of
the diffuse gamma--ray flux has an energy independent shape that
remains approximately equal to what what has been accurately measured
by the Fermi--LAT telescope in the GeV energy range,
indicating that the CR spatial distribution can only
undergo relatively minor changes over several decades of energy.
This conclusion does not support models where the cosmic rays have harder spectra
in the central part of the Galaxy.

Interestingly, the LHAASO measurements
indicate  that, for $E_\gamma \gtrsim 30$~TeV,
the ratio of the fluxes in the inner--Galaxy
and outer--Galaxy sky regions {\em decreases} with energy.
Statistical and systematic uncertainties are large, but this result,
if confirmed, would have great importance as it implies
that cosmic rays in the inner part of the Milky Way have
softer spectra than those in the outer part.
A possible explanation for this result is that the 
cosmic rays sources emit (on average) spectra with a position independent shape,
but the CR confinement volume increases with energy.
An alternative explanation is that a significant fraction of
cosmic rays in the multi--PeV energy range is of extragalactic origin.
The theoretical expectation is that, to a good approximation,
extragalactic particles have a uniform spatial distribution in the Milky Way volume,
or perhaps even a distribution that increases with distance
from the Galactic center. The existence of an extragalactic cosmic--ray component
that becomes gradually dominant in the energy range under study
could then explain the observations.

More in general, it should be stressed that the
study of the CR spatial distribution at very high energy
is an interesting method for determining whether a cosmic ray component
is Galactic or extragalactic.
In this context, one can note that some models
for the diffuse gamma and neutrino fluxes
include as a parameter the maximum energy of cosmic ray particles of Galactic origin.
For example the superscript (5 or 50) in the KRA$_\gamma^5$ and KRA$_\gamma^{50}$ models
\cite{IceCube:2023ame,ANTARES:2018nyb}
indicates the cutoff energy in PeV of the Galactic proton component.
However, a precise calculation of the
Galactic diffuse $\gamma$ and $\nu$ fluxes must also
include the modeling of the energy spectra
and spatial distribution of extragalactic cosmic rays,
because also these particles contribute to the interstellar emission 
when they interact with the Milky Way interstellar gas.
Including this contribution would reduce the differences between the results
obtained using the two templates.

As discussed above, the study of the energy dependence of the
angular distribution of the diffuse fluxes allows to determine
whether the spatial and energy distribution are factorized,
without the need to model the interstellar gas distribution.
On the other hand, to infer the properties of the CR spatial distribution,
(in a certain energy range)
some modeling of the interstellar gas distribution is necessary.
In this work we discuss how the angular distribution of the
gamma--ray diffuse flux measured by Fermi--LAT at $E_\gamma \approx 10$~GeV,
clearly indicates that the cosmic rays that generate this flux,
and have an energy per nucleon of order 100~GeV
(see discussion in Appendix~\ref{sec:appa})
are not uniformly distributed in the Milky Way volume,
but have a spatial distribution that, as it is natural to expect,
falls with the distance from the Galactic center.

This conclusion is based on the comparison 
of the angular distribution of the diffuse gamma--ray flux measured by
Fermi--LAT with what is expected for a uniform CR spatial distribution.
In the latter case the diffuse flux from the direction $\Omega$ is simply
proportional to the column density of interstellar gas 
along the same line of sight. This column density can be directly measured,
and the Planck satellite dust maps \cite{Planck:2016frx} provide an accurate
measurement.
The shapes of the two angular distributions
(of the gas column density and the diffuse flux)
are very similar, with differences $\lesssim \pm 50\%$
(see Fig.~\ref{fig:diffuse_long}), but the differences clearly
indicate that the CR density is larger in the inner part of the Galaxy. 

To infer quantitatively the spatial distribution of the CR density requires the
modeling of the 3D distribution of interstellar gas,
a much more difficult task than the direct measurement of the gas column density.
A preliminary study, based on the simple (cylindrically symmetric)
model of the interstellar gas density
described in \cite{Lipari:2018gzn}, shows that the Fermi--LAT observations are
consistent with a CR density that depends exponentially
on the (cylindrical coordinate) radius $r$ ($\propto e^{-\lambda_{\rm cr} r}$)
with a length scale $\lambda_{\rm cr}^{-1}$ of order 5--6~kpc.

Currently the measurements of the diffuse gamma--ray and netrino fluxes
have large errors, and attempts to interpret these results
(such as this work) cannot yet reach robust conclusions. However, 
future observations have the potential to clarify many of
the questions discussed here. 
To achieve these goals, it is desirable to
obtain measurements that are sensitive to the properties of cosmic rays 
in as large a volume of the Milky Way as possible. This requires
measuring the diffuse $\gamma$ and $\nu$ spectra along lines of sight
that traverse the Galactic disk with small latitude $|b|$,
and those that pass near the Galactic center are the most interesting.
These lines of sight are also those that have a high probability
of passing close to point--like or quasi--point--like sources,
and for this reason most of these lines have 
been excluded (by masking known sources) in the LHAASO observations,
resulting in a significant reduction of sensitivity.
Including these lines of sight in future studies of the gamma--ray and neutrino
fluxes is very desirable. However, interpreting the data will 
require separating the component generated in (or near) the sources
from the component due to interstellar emission.
This is going to be a difficult, but very important task,
that has also the merit to explore the possibility
that the CR spatial distribution will not only have
a large scale smooth gradient, but also contain 
higher density ``bubbles'' around (at least some of) the sources.

\vspace{0.35 cm}
\noindent {\bf Acknowledgments} \\
We are grateful to Cao Zhen, Felix Aharonian,
Dmitriy Khangoulian, You Zhi Yong, Xi Shao Qian, Shoushan Zhang, Zha Min,
Pedro De La Torre Luque, Daniele Gaggero, Dario Grasso and Antonio Marinelli  
for useful discussions. \\
PL  acknowledges support from the Sichuan Science and Technology Department
under grant number 2024JDHJ0001.

\vspace{1 cm}

\appendix
\section{Relation between the gamma--ray and neutrino energy
 and the interacting nucleon energy}
\label{sec:appa}

The emission of gamma--rays or neutrinos of a certain energy $E_{\gamma, \nu}$
is generated by the interactions
of primary nucleons (including those that are  bound in a nucleus)
with a rather wide range of energies.
The contribution of CR of energy $E_0$
to the emission of gamma--rays with energy $E_\gamma$ from the point $\vec{x}$ is:
\begin{equation}
 \frac{dq_\gamma (E_0; E_\gamma,\vec{x})}{d\ln E_0} \propto 
 E_0 \; N_{\rm cr} (E_0, \vec{x}) ~\frac{dN_\gamma}{dE_\gamma} (E_\gamma, E_0)
\label{eq:dqgamma}
\end{equation}
where $N_{\rm cr} (E_0, \vec{x})$ the
density of primary particles at the point $\vec{x}$.
A similar expression, with obvious substitutions is valid for neutrinos.

\begin{figure}[bt]
\begin{center}
\includegraphics[width=12.9cm]{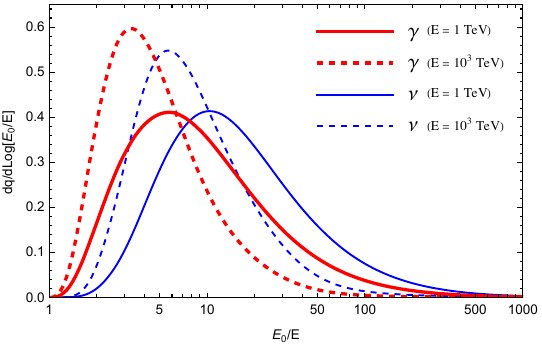}
\end{center}
\caption {\footnotesize
 Examples of the (normalized)
 distributions $dq_{\gamma (\nu)} (E_{\gamma (\nu)}, E_0)/d\log E_0$
 of the gamma--ray (neutrino) interstellar emission
 at energy $E_{\gamma(\nu)} = 1$~TeV and~$10^{3}$~TeV,
 produced in the interactions of primary particles with energy per nucleon $E_0$.
 The distributions are calculated under the assumption that the cosmic ray spectra
 at the emission point are described by the GST model
 \cite{Gaisser:2013bla} and by using the Sibyll--2.1 code \cite{Ahn:2009wx}
 to model hadronic interactions.
 \label{fig:response_gam_nu}}
\end{figure}

Some examples of the distributions of primary nucleon energy that contribute
to the emission, calculated by  assuming  that the primary cosmic ray spectra
have the shape of the GST model \cite{Gaisser:2013bla}
and by using  the Sibyll--2.1 code \cite{Ahn:2009wx}
to model hadronic interactions, 
are shown in Fig.~\ref{fig:response_gam_nu}.  The parameters that characterize
the  shapes  of these distributions are listed in Table~\ref{tab:response}.
\begin{table}[hbt]
 \caption{\footnotesize
   Quantities that describe the distributions
   of primary nucleon energy
   (shown in Fig.~\ref{fig:response_gam_nu})
 that generate the gamma--ray and neutrino interstellar emission
 at $E = 1$~TeV and $E = 10^3$~TeV.
  $E_0^{\rm (peak)}$ is the energy where
 the distribution $dq_{\gamma(\nu)} /d\log E_0$ has its maximum, 
 and $E_0^{\rm (median)}$ is the median of the distribution.
 The interval of primary nucleon energy [$E_0^{\rm (low)}$, $E_0^{\rm (high)}$] generates
 50\% of the emission, while the
 intervals [$E_\gamma$, $E_0^{\rm (low)}$]
 and [$E_0^{\rm (high)}$, $\infty$] generate 25\% of the emission.
\label{tab:response}}
 \renewcommand{\arraystretch}{1.4}

 \vspace{0.20cm}
 \begin{tabular}{ | l || c | c | c | c |}
 \hline
 ~~~
& $E_0^{\rm (peak)}/E$ 
& $E_0^{\rm (median)}/E$ 
& $E_0^{\rm (low)}/E$ 
& $E_0^{\rm (high)}/E$ \\
 \hline
 ~ $\gamma$ ~($E = 1$~TeV) ~
 & 5.9	& 7.5	& 4.0	& 15.9 \\
 ~ $\nu$ ~($E = 1$~TeV) ~
& 10.2	& 13.2	& 7.1	& 27.9 \\
 ~ $\gamma$ ~($E = 10^3$~TeV) ~
& 3.4	& 4.1	& 2.7	& 7.1 \\
 ~ $\nu$ ~($E = 10^3$~TeV) ~
& 5.6	& 7.0	& 4.4	& 12.5 \\
\hline
\end{tabular}
\end{table}

The range of nucleon energy that contribute to gamma--rays 
of energy $E_{\gamma}$ extends for approximately a decade centered
at a value $E_0/E_{\gamma} \simeq 4$--7.
For the  production of neutrinos at energy $E_\nu$ 
the contributing range of nucleon energies has
approximately the same width, but is higher by a factor of order two,
reflecting the fact that charged and neutral pions are
produced in inelastic hadronic collisions with spectra of
similar shape, but the three neutrinos produced (together with an
$e^\pm$) in the chain decay of a $\pi^\pm$
carry approximately one--quarter of the parent energy each,
while the two gamma--rays generated in $\pi^0$ decay
have on average half of its energy.

The relation between the gamma--ray (neutrino) energy and
the interacting nucleon energy obviously depends on the
shape of the primary CR spectra in the relevant range.
In the example considered the
average $ \langle E_0\rangle /E_{\gamma (\nu)}$ is significantly smaller
for $E_{\gamma (\nu)} = 1$~PeV, because in our calculation
the CR spectra in the multi--PeV range
undergo the well known ``knee'' softening.

\clearpage

\end{document}